\documentclass[aps,prd,twocolumn,groupedaddress,amssymb,eqsecnum,showpacs,psfig]{revtex4}
\usepackage{graphicx}
\usepackage{bm}
\usepackage{dcolumn}
\usepackage{amsmath}

\begin{document}

\newcommand{\newc}{\newcommand}

\newc{\be}{\begin{equation}}
\newc{\ee}{\end{equation}}
\newc{\ba}{\begin{eqnarray}}
\newc{\ea}{\end{eqnarray}}
\newc{\bea}{\begin{eqnarray*}}
\newc{\eea}{\end{eqnarray*}}
\newc{\D}{\partial}
\newc{\ie}{{\it i.e.} }
\newc{\eg}{{\it e.g.} }
\newc{\etc}{{\it etc.} }
\newc{\etal}{{\it et al.}}
\newcommand{\nn}{\nonumber}

\newc{\ra}{\rightarrow}
\newc{\lra}{\leftrightarrow}
\newc{\no}{Nielsen-Olesen }
\newc{\lsim}{\buildrel{<}\over{\sim}}
\newc{\gsim}{\buildrel{>}\over{\sim}}
\title{Equation of State of Oscillating Brans-Dicke Scalar and Extra Dimensions}
\author{L. Perivolaropoulos}
\email{leandros@physics.uoi.gr} \affiliation{Department of
Physics, University of Ioannina, Greece}
\date{\today}

\begin{abstract}
We consider a Brans-Dicke scalar field stabilized by a general
power law potential with power index $n$ at a finite equilibrium
value. Redshifting matter induces oscillations of the scalar field
around its equilibrium due to the scalar field coupling to the
trace of the energy momentum tensor. If the stabilizing potential
is sufficiently steep these high frequency oscillations are
consistent with observational and experimental constraints for
arbitrary value of the Brans-Dicke parameter $\omega$. We study
analytically and numerically the equation of state of these high
frequency oscillations in terms of the parameters $\omega$ and $n$
and find the corresponding evolution of the universe scale factor.
We find that the equation of state parameter can be negative and
less than -1 but it is not related to the evolution of the scale
factor in the usual way. Nevertheless, accelerating expansion is
found for a certain parameter range. Our analysis applies also to
oscillations of the size of extra dimensions (the radion field)
around an equilibrium value. This duality between self-coupled
Brans-Dicke and radion dynamics is applicable for $\omega= -1 +
1/D$ where D is the number of extra dimensions.

\end{abstract}

\maketitle

\section{Introduction}
Observations of the magnitude-redshift relation of distant Type Ia
supernovae\cite{Riess:1998cb} and CMB
measurements\cite{deBernardis:2001xk} have indicated that the
universe is undergoing a period of accelerated expansion.
Attributing this acceleration to a cosmological constant is not
completely satisfactory (see e.g. \cite{Sahni:1999gb}) because
fine tuned initial conditions are required in order to solve the
coincidence problem (why the vacuum energy is dominating the
energy density right now). Thus, acceleration is attributed to the
influence of a redshifting non-luminus form of energy with
negative pressure called by many authors `dark energy' or
`quintessence'\cite{Peebles:1987ek}. The origin of this form of
energy remains unknown. Several types of scalar
fields\cite{Peebles:1987ek} have been proposed as sources of dark
energy including the dilaton, the inflaton, supersymmetric
partners of fermions, Brans-Dicke (hereafter BD)
scalars\cite{Brans:sx}, etc.

The dynamical evolution of these scalars is determined by
specially designed potentials so that the scalar field energy
density tracks the radiation-matter component at early times while
at late times it dominates and leads to acceleration. The main
drawback of these models is that there is usually no physical
motivation for the proposed scalar fields  and potentials.

Cosmological theories with extra dimensions generically contain a
scalar field of geometrical origin which describes the size of
extra dimensions. This is known as the radion field. These
theories have the potential of providing a physically motivated
solution to the hierarchy problem by postulating that the
fundamental Planck mass $M_*$ is close to the $TeV$ scale
\cite{Antoniadis:1990ew,pap.Dim,Randall:1999vf}. This is possible
in theories with extra dimensions \cite{pap.Dim} because Gauss's
law relates the Planck scales of the $4+D$-dimensional theory
$M_*$ and the long distance $4$-dimensional theory $M_{pl}$ by \be
M_{pl}^2 = b_0^D M_*^{D+2} \ee where $b_0$ is the present
stabilized size of the extra dimensions.

According to observations, the internal space should be static or
nearly static at least from the time of primordial
nucleosynthesis\cite{GZ(CQG2)}. This means that at the present
evolutionary stage of the Universe only small fluctuations over
stabilized or slowly varying compactification scales (conformal
scales/geometrical moduli) are possible. Thus the size of the
extra dimensions $b(t)$ (the radion field) is usually assumed to
be stabilized to its present value $b_0$ by a `radion stabilizing
potential' $V(b)$. Observational constrains require this
stabilization and theoretical models\cite{Goldberger:1999uk} have
been proposed to justify it. Within the framework of
multidimensional cosmological models the role of the size of extra
dimensions as a stabilized scalar field with possible excited
states was investigated in \cite{GZ1,GZ,GZ(PRD2)} where they were
called gravitational excitons.

In a recent paper\cite{Perivolaropoulos:2002pn} we demonstrated
that a generic feature of these theories is the existence of
radion oscillations (induced by redshifting matter density) around
the equilibrium value $b_0$ at the minimum of the stabilizing
potential. It was also shown that these oscillations can lead to
cosmological periods of accelerating expansion of the universe.
However the radion mass required for such extremely low frequency
oscillations was too small (at the classical level) to be
consistent with fifth force experimental
constraints\cite{Will:2001mx, Steinhardt:1994vs}. In the same
paper\cite{Perivolaropoulos:2002pn} it was also demonstrated that
the dynamics of the radion field are formally equivalent with the
dynamics of a massive BD scalar with $\omega=-1+1/D$ where $D$ is
the number of extra dimensions (see also \cite{Cline:2002mw,
Levin:yw}).

Quintessence models based on BD scalars (extended quintessence)
have the theoretically appealing property that the same scalar
field that participates in the gravity sector is enhanced by a
potential\cite{Santos:1996jc} to play the role of quintessence.
Even though no oscillating BD scalars have been studied so far,
there is an extensive literature on extended
quintessence\cite{Torres:2002pe,Faraoni:2001tq,Perrotta:1999am}.

Here we focus on the cosmological evolution of high frequency
oscillations\cite{Accetta:yb, Steinhardt:1994vs} of a BD
scalar\cite{Brans:sx} (or equivalently a radion field) stabilized
by an arbitrary power law potential. For large enough frequencies
$\nu \gsim 10^{12}Hz$ these oscillations are consistent with fifth
force constraints\cite{Accetta:yb, Steinhardt:1994vs} and they can
play the role of dark matter and/or dark energy for a wide range
of parameters.

The structure of this paper is the following: In section II we
review the cosmological evolution of an oscillating minimally
coupled scalar field. In section III we  demonstrate the
equivalence between radion dynamics in $D$ extra dimensions and
massive BD scalar $\phi$ with $\omega=-1+1/D$, $\phi=b^{2D}$ and
focus on the dynamics of an oscillating BD scalar. We derive a
virial theorem that connects the kinetic and potential energies of
the oscillating field in terms of the parameters $\omega$ and $n$.
This theorem is verified by numerically solving the dynamical
cosmological field equations for various parameter values. We then
use this virial theorem to find the equation of state that
connects the effective pressure with the effective energy density
of the oscillating field. In section IV we use the equation of
state to find the redshift rate of the effective energy density
and the evolution of the cosmological scale factor. We find that
the energy redshift and the scale factor evolution are independent
of $\omega$ and depend only on the stabilizing potential power
index $n$. The validity of these results is demonstrated
numerically. Finally in section V we conclude and discuss possible
extensions and open issues related to this work. In what follows
we set $8\pi G=1$.

\section{Minimally Coupled Oscillating Scalar}
As a warm up exercise we evaluate the equation of state for a
minimally coupled coherently oscillating scalar in an expanding
background. This analysis was originally done in Ref.
\cite{Turner:1983he} but we briefly review it here for
completeness and for comparison of the results with the BD scalar
oscillations analysis performed in the following sections.

Consider a minimally coupled scalar field whose dynamics is
determined by the Lagrangian density \be {\cal L}={1\over 2}
\partial_\mu \phi \partial ^\mu \phi - V(\phi) \ee
Assuming a Robertson-Walker  metric we obtain the field equation
\be \label{fieldeq}\ddot \phi + 3 H \dot \phi  = -
\frac{dV}{d\phi},\ee where $H={{\dot a}\over a}$. From the energy
momentum tensor we obtain the energy density $\rho$ and the
pressure $p$ as \ba \label{rhodef1} \rho &=&{1\over 2} {\dot \phi}^2 + V(\phi) \\
\label{pdef1} p&=&{1\over 2} {\dot \phi}^2 - V(\phi) \ea Using
equations (\ref{fieldeq}) and (\ref{rhodef1}) we obtain \be
\label{rhoev1} {d\over{dt}} \rho=-3H \dot \phi^2 \ee We now
consider a potential of the form \be \label{pot1} V(\phi) =
\lambda |\phi |^n \ee where $n>0$ and focus on oscillations of the
scalar field around its minimum. We also define the mean value of
a time dependent quantity $q(t)$ as \be {\bar{q}}\equiv {1\over T}
\int_0^T dt \; q(t) \ee where T is the period of one oscillation.
The mean equation of state of the oscillating field is \be
\label{eqst1}\bar{p} = w \bar{\rho} \ee  We now evaluate the
equation of state parameter $w$ in terms of $n$.

Using equations (\ref{rhodef1}), (\ref{pdef1}), (\ref{rhoev1}) and
(\ref{eqst1}) we obtain \be \label{drhodt}{{d\bar{\rho}}\over
{dt}} =- 3H \bar{\dot{\phi^2}} = -3H\gamma \bar{\rho}= -3{{\dot
a}\over a} \gamma \bar{\rho} \ee where $\gamma=w+1$ and we have
assumed $T \ll H^{-1}$. Thus, on timescales $\Delta t \ll H^{-1}$
we have ${{\Delta \bar{\rho}}\over {\bar{\rho}}}\ll 1$ and we can
set \be \label{rhocons1}  \rho ={1\over 2} {\dot \phi}^2 + V(\phi)
\simeq V(\phi_{max}) \equiv V_{max} = \bar{\rho} \ee where
$\phi_{max}$ is the amplitude of the $\phi$ oscillations and
$V_{max}$ is the corresponding maximum potential energy. We now
have \be \label{gamma1} \gamma = {{\bar{p} +\bar{\rho}}\over
{\bar{\rho}}} = {{\bar{\dot{\phi^2}}}\over {\bar{\rho}}}=
{\sqrt{2}\over T_{max}}\int_0^{T_{max}}{{d\phi}\over
V_{max}^{1/2}} (1-V/V_{max})^{1/2} \ee We also use equation
(\ref{rhocons1}) to find the time $T_{max}={T\over 4}$ it takes
the field to go from 0 to $\phi_{max}$ \be \label{tmax}
 T_{max}={1\over
{\sqrt{2}\; V_{max}^{1/2}}} \int_0^{\phi_{max}} d\phi
(1-V/V_{max})^{-1/2} \ee Using equations (\ref{gamma1}) and
(\ref{tmax}) we find \be \label{intrat1} \gamma=
2{{\int_0^{\phi_{max}} d\phi (1-V/V_{max})^{1/2}}\over
{\int_0^{\phi_{max}} d\phi (1-V/V_{max})^{-1/2}}} =
{{2n}\over{n+2}}\ee where we made use of (\ref{pot1}).

The energy density redshift is now obtained from equation
(\ref{drhodt}) as \be \label{rhoa} \rho = \rho_0 \left({a\over
a_0} \right)^{-3\gamma} \ee Using this equation and the Friedman
equation \be {{\dot a}\over a}^2={\rho \over 3} \ee
 we obtain \be \label{aevol1} a\sim t^{2/3\gamma} = t^{{n+2}\over {3n}} \ee
and accelerated expansion is obtained in the range $0<n<1$. The
cosmological effects of this accelerated expansion for an
oscillating minimally coupled scalar have been discussed in Ref.
\cite{Sahni:1999qe}. The scalar field evolution for power index
$n$ in the above range is well defined even though for $n<1$,
$V'(\phi)$ is weakly singular at $\phi=0$. $V(\phi )$ can be made
mathematically more appealing by the field redefinition $\phi
\rightarrow (\phi^2 + \phi_c^2)^{1/2}$, $\phi_c \rightarrow 0$
which does not change the above results\cite{Sahni:1999qe}. In the
next section we will show that even though the equation of state
is different for an oscillating BD scalar, accelerated expansion
is obtained for the same range of the power index $n$.

\section{Virial Theorem  and Equation of State for Oscillating Radion and Brans-Dicke Scalar}

In this section we demonstrate the equivalence between BD and
Radion dynamics (for specific values of the BD parameter $\omega$)
and then focus on the BD theory to derive a virial theorem and the
equation of state of the scalar field oscillations.

Consider the BD action
\begin{equation} \label{bdact}
{\cal S}=\frac{1}{2} \int d^4x \sqrt{-g}\left[ \phi
R-{\omega\over{\phi}}\partial_\mu \phi
\partial^\mu\phi - 2 V(\phi)\right]+ L_{fluid}
\end{equation}
where $L_{fluid}$ describes matter and/or radiation. We assume a
flat ($k=0$) Friedmann-Robertson-Walker model whose metric reads
\begin{equation} \label{rw}
ds^2=-dt^2+a^2(t)[dr^2+r^2 d\theta^2+r^2\sin^2\theta d\phi^2]
\end{equation}
The combination of equations (\ref{bdact}) and (\ref{rw}) leads to
the dynamical equations for $\phi$ and $a$
\begin{equation} \label{bd1}
 {\dot a^2\over{a^2}}+{\dot
a\over{a}}{\dot\phi\over{\phi}}-{\omega\over{6}}
{\dot\phi^2\over{\phi^2}}-{ V\over{3\phi}}={\rho\over{3\phi}},
\end{equation}
\begin{equation} \label{bd2}
 2{\ddot{a}\over{a}}+{\dot a^2\over{a^2}}+{\ddot
{\phi}\over{\phi}}+2{\dot
a\over{a}}{\dot\phi\over{\phi}}+{\omega\over{2}}
{\dot\phi^2\over{\phi^2}}-{ V\over{\phi}}=-{ p\over{\phi}},
\end{equation}
\be \label{bd3}  {\ddot{\phi}}+3{\dot
a\over{a}}{\dot\phi}={(\rho-3p)\over{2\omega+3}}+
{2\over{2\omega+3}}\left[2V-\phi{dV\over{d\phi}}\right]  \ee
From the field equations we can read the {\it effective energy and
pressure} (see e.g. \cite{Sen:2000vj,Torres:2002pe})
 for the field which end up being,
\begin{equation} \label{rhoeq}
 \rho_\phi=
{\omega\over{2}}~{\dot\phi^2\over{\phi}}+V - 3 {\dot
a\over{a}}\dot\phi
\end{equation}
and
\begin{equation} \label{peq}
 p_\phi={\omega\over{2}}~{\dot\phi^2\over{\phi}}-{
V}+\ddot\phi +2{\dot a\over{a}}\dot\phi
\end{equation}
The BD theory is interesting in its own right as a generalization
of Einstein gravity. In addition however many physically motivated
theories reduce to BD gravity for specific values of $\omega$. For
example the superstring dilaton is akin to a BD theory with
$\omega=-1$. Also Einstein gravity generalized to 4+D dimensions
reduces to an effective 4-dimensional BD theory with
$\omega=-1+1/D$. This can be demonstrated as follows:

We consider a 3-brane embedded in a toroidally compact ($4+D$)
dimensional space-time $R^1\times R^3\times T^D$ with the $D$
extra spatial dimensions (the bulk) stabilized at a size $b_0$.
The total action may be written as \be S_{tot}= S_{bulk} +
S_{brane} \ee where \be S_{bulk} =
 \int d^{4+D}x \sqrt{-g{^{(4+D)}}} \biggl( \frac{1}{16\pi
\overline{G}}{\cal {R}}[g] + {\cal L_{\rm bulk}} \biggr)
\label{action in 4+D} \ee and \be S_{brane} =
 \int d^{4}x \sqrt{-g{^{(4)}}}\; {\cal L_{\rm brane}}
\label{action in brane} \ee  In (\ref{action in 4+D}) ${\cal
L_{\rm bulk}}$ is the Lagrangian of the bulk fields apart from the
graviton. These fields give rise to the stabilizing potential $V$
discussed below. The background metric for the a toroidally
compact ($4+D$) dimensional space-time which is consistent with
the symmetries of the brane-bulk system may be written as \be
\label{eq101}
g_{MN}=diag[1,-a^2(t)\tilde{g}_{ij},-b^2(t)\tilde{g}_{mn}] \ee
where $M,N$ run from 0 to $D+3$; $ i , j $ run from $1$ to 3 and
$m,n$ run from 4 to $D+3$. Also $a(t)$ is the scale of the
non-compact 3-dimensional flat space (the scale factor of the
universe) and $b(t)$ is the radius of the compactified toroidal
space (the radion field).

We assume that matter on the brane ($S_{brane}$) is represented by
a classically conserved perfect fluid energy momentum tensor
$T_{\mu \nu} = (\rho + p)u_\mu u_\nu + p g_{\mu \nu}$ ($\mu, \;
\nu = 0,...,3$) with $\nabla_\mu T^{\mu \nu}=0$ where $u_\mu$ is a
future-oriented time-like vector $u^\mu = (1,\vec{0}_{3+D})$ in
the basis (\ref{eq101}) and $\rho$ is the energy density of the
brane matter. Also $p=w\rho$ is the corresponding pressure. We
also include a radion stabilizing potential \be V(b)\equiv b(t)^D
\int d^{4+D}x\; {\cal L_{\rm bulk}} \ee induced by some
non-specified bulk dynamics ${\cal L_{\rm
bulk}}$\cite{Goldberger:1999uk}. For example a uniform bulk
cosmological constant is represented by $V=b^D \Lambda$ while a
Casimir type potential would be $V(b)=k b^{-4}$
\cite{Gardner:2001fz}. We assume that bulk dynamics lead to a
potential $V(b)$ that stabilizes $b$ at
%
%
\be b_{0}=\left({M_{Pl}^2 \over M_{*}^{D+2}}\right)^{1/D} \ee with
a vanishing cosmological constant.  The equations of motion for
the coupled $(a(t), b(t))$ system can be written
as\cite{Arkani-Hamed:1999gq}
\begin{widetext}
\ba \label{eqrad} &&6 \frac{\dot a^2}{a^2} + D(D-1)\frac{\dot
b^2}{b^2} + 6D \frac{\dot
a}{a} \frac{\dot b}{b} = 2\frac{V + \rho}{b^D} \nonumber \\
&&\frac{\ddot b}{b} + (D-1) \frac{\dot b^2}{b^2} + 3 \frac{\dot
a}{a}\frac{\dot b}{b} =2 \frac{1}{ b^D} \left(\frac{2 V}{D+2} -
\frac{b}{D(D+2)} \frac{\partial V}{\partial b} +
\frac{\rho-3p}{2(D+2)} \right)
\\ &&\frac{\ddot a}{a} + 2\frac{\dot a^2}{a^2} + D \frac{\dot
a}{a}\frac{\dot b}{b} = 2\frac{1}{b^D} \left( \frac{b}{2(D+2)}
\frac{\partial V}{\partial b} - \frac{D-2}{2(D+2)} V + \frac{\rho
+ (D-1) p}{2(D+2)} \right) \nonumber \ea \end{widetext}

In the picture of Refs. \cite{pap.Dim} (see also
\cite{Arkani-Hamed:1999gq}) (ADD) and \cite{Randall:1999vf} (RS),
the matter-radiation energy density is assumed to be localized on
the brane corresponding to the $a(t)$ scale factor. This localized
energy density in general distorts the geometry of the
compactified $D$-dimensional space (the bulk), but as far as the
overall properties and the evolution of the radion are concerned,
it is correct to treat the energy density on the wall as just
being averaged over the whole space as done on the RHS of these
equations. This assumption is consistent with the results of
references \cite{Csaki:1999mp,Csaki:1999jh} where it was shown
that the coupling of the radion field to the energy momentum
tensor is given generically through the trace $(\rho-3p)$ plus
terms involving the stabilizing potential $V(b)$.

Equations (\ref{eqrad}) have been analyzed in the context of
inflation (away from the stabilization point $b_{0}$) in
\cite{Arkani-Hamed:1999gq} (ADD model).

Similar equations \cite{Csaki:1999mp} arise in the context of
Randall-Sundrum (RS) models \cite{Randall:1999vf}, where the
hierarchy problem is solved using extra dimensions of smaller
sizes at the expense of introducing a non-flat background metric
along the extra coordinates and a pair of branes whose distance
$b(t)$ is stabilized at $b_{0}$ by the potential $V(b)$.

From equations (\ref{eqrad}) it is clear that radiation
$(p_{rad}={1\over3} \rho_{rad})$ has no effect on the dynamics of
the radion. This is not the case however for matter. Redshifting
matter plays the role of a driving force and can induce radion
oscillations during both the matter and radiation eras. These
oscillations backreact on the scale factor $a(t)$ through the
effective Friedman equations (\ref{eqrad}) and can affect the
expansion rate.

It is interesting to compare the form of the cosmological
equations (\ref{eqrad}) with the non-conventional cosmology on the
brane of Ref. \cite{Binetruy:1999ut} which has shown that terms of
$O(\rho^2)$ appear on the Friedman equation due to the extra
dimensions. These terms can also be obtained from the system
(\ref{eqrad}) in the limit $({{\dot b}\over b})^2 \ll \rho$. In
that limit, it is a good approximation to set ${\dot b} =0$ in the
second of equations (\ref{eqrad}). For a stabilizing potential of
the form $ V \sim (b-b_0)^n $ we thus obtain $V\sim
\rho^{{n\over{n-1}}}$ and for a standard parabolic potential
($n=2$) we obtain an extra $\rho^2$ contribution due to the
potential in the first equation of the system (\ref{eqrad})
similar to the $\rho^2$ terms of Ref. \cite{Binetruy:1999ut}. For
a generalized stabilizing potential the extra term in the Friedman
equation is of the form $\rho^\alpha$ ($\alpha ={n\over{n-1}}$)
and can provide a physical motivation for Cardassian
expansion\cite{Freese:2002sq}. In what follows however we will not
use the above approximation which ignores the effects of radion
oscillations and we will focus on the cosmological expansion
induced by these oscillations for a generalized stabilizing
potential.

It is straightforward to show that we can obtain the  set of
dynamical equations (\ref{eqrad}) for the radion by setting to the
corresponding BD set of equations (\ref{bd1}), (\ref{bd2}),
(\ref{bd3}) \bea \phi &=& b^D
\\ \omega & =& -1+{1\over D}\eea Thus the radion dynamics is
equivalent to BD dynamics for  particular values of $\omega$. Even
though our study will be based on the general BD field equations
(\ref{bd1}), (\ref{bd2}) and (\ref{bd3}) the main physical
application of our results will be $4+D$ dimensional Einstein
gravity for which $\omega=-1+1/D$. Even though these values of
$\omega$ are of $O(1)$ they are consistent with experimental and
observational constraints because of the presence of the potential
$V$. This stabilizing potential can confine the effects of any
modifications of Newton's law to scales smaller than $1mm$
\cite{Accetta:yb,Steinhardt:1994vs,Perivolaropoulos:2002pn}. On
these scales there are no firm experimental constraints on the
form of the gravitational potential\cite{Will:2001mx}.

We now consider the BD field equations (\ref{bd1}), (\ref{bd2}),
(\ref{bd3}) with a potential $V(\phi)$ having a minimum at
$\phi_0$. It is clear from equation (\ref{bd3}) that even if we
consider an initially static field $\phi=\phi_0$ the redshifting
matter will induce small oscillations around the equilibrium
position  (approximately $\phi_0$) by slowly shifting the
equilibrium position of the field $\phi$. We will thus study the
equation of state of the energy density of these small
oscillations assuming a potential of the form \be V(\phi)=\lambda
|\phi - \phi_0|^n \equiv \lambda |\delta\phi|^n \ee  Since not all
three equations of the system (\ref{bd1}), (\ref{bd2}) and
(\ref{bd3}) are independent (because of the Bianchi identities) we
focus on (\ref{bd1}) and (\ref{bd3}) and write them in
dimensionless form as
\begin{equation} \label{bdd1}
 {\dot a^2\over{a^2}}+{\dot
a\over{a}}{\dot\phi\over{\phi}}-{\omega\over{6}}
{\dot\phi^2\over{\phi^2}}={ 1\over{\phi}}({1\over {a^3}}+
\bar{\rho_v}|\phi -1|^n)
\end{equation}
\be \label{bdd3}  {\ddot{\phi}}+3{\dot
a\over{a}}{\dot\phi}={3\over{(2\omega +3)}}({1\over {a^3}}+ 4
\bar{\rho_v}|\phi -1|^n\mp 2 n \phi \bar{\rho_v}|\phi -1|^{n-1})
\ee where we have assumed the presence of matter redshifting like
\be \rho_m(t)=\rho_{0m} ({{a_0}\over {a(t)}})^3 \ee (radiation has
no effect on the evolution of $\phi$ since $\rho_r - 3 p_r =0$).
We have also set $a_0\equiv a(t_p)=1$ where $t_p$ is the present
time) and \ba \label{tresc} {\rho_{0m} \over 3} t &\longrightarrow  & t \\
{\phi \over \phi_0}  &\longrightarrow  & \phi \\
\label{rhoresc}{{\lambda \phi_0^n} \over \rho_{0m}}
&\longrightarrow & \bar{\rho_v} \ea  The $-$ sign ($+$ sign) in
equation (\ref{bdd3})  is valid when $\phi > 1$ ($\phi < 1$). As
in the case of the minimally coupled scalar of section II $\phi$
is periodic on timescales $\Delta t << H^{-1}$. With no loss of
generality we may therefore focus on the half period $T\over 2$
for which $\delta \phi \equiv \phi -1 >0$. We also assume $\delta
\phi \ll 1$. Thus equation (\ref{bdd3}) is written as \be
\label{bddd3} \ddot{\delta \phi}  + {{6 n \bar{\rho_v}}\over {2
\omega + 3}} \delta \phi^{n-1}={3 \over {2 \omega +3}} {1\over
a^3} - 3{\dot a\over{a}}\dot{\delta\phi} \ee This implies that \be
\label{bdecon} {d\over {dt}} ({{\dot{\delta \phi^2}}\over 2} +{{6
\bar{\rho_v}}\over {2 \omega + 3}} \delta \phi^n - {3 \over {2
\omega +3}} {1\over a^3} \delta \phi)= - 3{\dot
a\over{a}}\dot{\delta\phi^2}  \ee Therefore the effective
potential that determines the dynamics of the oscillating BD
scalar is \be \label{veff0} V^{eff}(\phi) ={{6 \bar{\rho_v}}\over
{2 \omega + 3}} \delta \phi^n - {3 \over {2 \omega +3}} {1\over
a^3} \delta \phi \ee For $n>1$ the effective potential is not
minimized at $\delta \phi=0$ but at the time-dependent value \be
\label{dfmin} \delta\phi_{eql}= ({3 \over {6 n \bar{\rho_v}}}
{1\over a^3})^{1\over {n-1}} \ee If the amplitude of oscillations
$\delta \phi_{max}$ is much larger than the shift of equilibrium
point due to matter redshift then we may ignore the shift and
assume that the oscillations take place around $\delta \phi=0$. In
the opposite case when $\delta \phi_{eql} \gg \delta \phi_{max}$
we may ignore the oscillations and assume that $\delta \phi \simeq
\delta \phi_{eql}$ at all times. In the later case the energy
density of the BD scalar is dominated by potential energy V which
redshifts like \be V(\phi)\sim a^{-3n/(n-1)} \ee and is therefore
subdominant compared to matter at late times. Thus in what follows
we focus on the former case assuming \be \label{oscdom} \delta
\phi_{max} \gg \delta \phi_{eql}\ee It will be shown that for
$n<4$ this condition is an attractor in the sense that even if it
is not realized at early times it becomes true at late times i.e.
$\delta \phi_{max}$ redshifts slower than $\delta \phi_{eql}$.
Thus since the oscillations of $\delta \phi$ take place
symmetrically around $\delta \phi = 0$ and since the period  and
amplitude are approximately constant on timescales much less than
$H^{-1}$ the mean value of $\delta \phi$ (and of $\dot{\delta
\phi}$) vanishes and equation (\ref{bdecon}) implies \be
\label{bdecon1} {{\bar{\dot{\delta \phi^2}}}\over 2} + {{6
\bar{\rho_v}}\over {2 \omega + 3}} \bar{\delta \phi^n}   \simeq
{{6 \bar{\rho_v}}\over {2 \omega + 3}} \delta \phi_{max}^n \equiv
V_{max}^{eff} \ee Proceeding in similar way as in the previous
section (equations (\ref{drhodt}) - (\ref{intrat1})) we have \be
\label{intrat2} \bar{\dot{\delta\phi^2}}= 2
V_{max}^{eff}{{\int_0^{\delta\phi_{max}} d\delta\phi
(1-V^{eff}/V_{max}^{eff})^{1/2}}\over {\int_0^{\delta\phi_{max}}
d\delta\phi (1-V^{eff}/V_{max}^{eff})^{-1/2}}} = V_{max}^{eff}
{{2n}\over{n+2}}\ee Using now this result and equations
(\ref{veff0}), (\ref{bdecon1}) we find \be \label{veff1}
\bar{V}^{eff} = V_{max}^{eff}(1 - {n\over {n+2}}) = {2\over {n+2}}
V_{max}^{eff} \ee and \be \label{virial}
{{\bar{\dot{\delta\phi^2}}} \over \bar{V^{eff}}}=n \Rightarrow
{{\bar{\dot{\delta\phi^2}}} \over
{\bar{\rho_v}\bar{\delta\phi^n}}}= {{6n}\over {2\omega + 3}} \ee
This result is a virial theorem for BD scalar oscillations
connecting the mean kinetic and potential energies. We have
numerically confirmed the validity of this theorem by solving the
full system (\ref{bdd1}), (\ref{bdd3}) for $\bar{\rho_v}\gg 1$
(high frequency oscillations) from $t_i = {1\over
\sqrt{\bar{\rho_v}}}$ to $t_0 = 1$. We used initial
conditions \ba \dot{\delta \phi}(t_i) &=& 0 \\
\delta \phi(t_i)&=& \delta \phi_{max} \ll 1 \\
a(t_i) &=& (2/3)^{-2/3} t_i^{-2/3} \ea
\begin{figure}
\centering
\includegraphics[height=8cm,angle=-90]{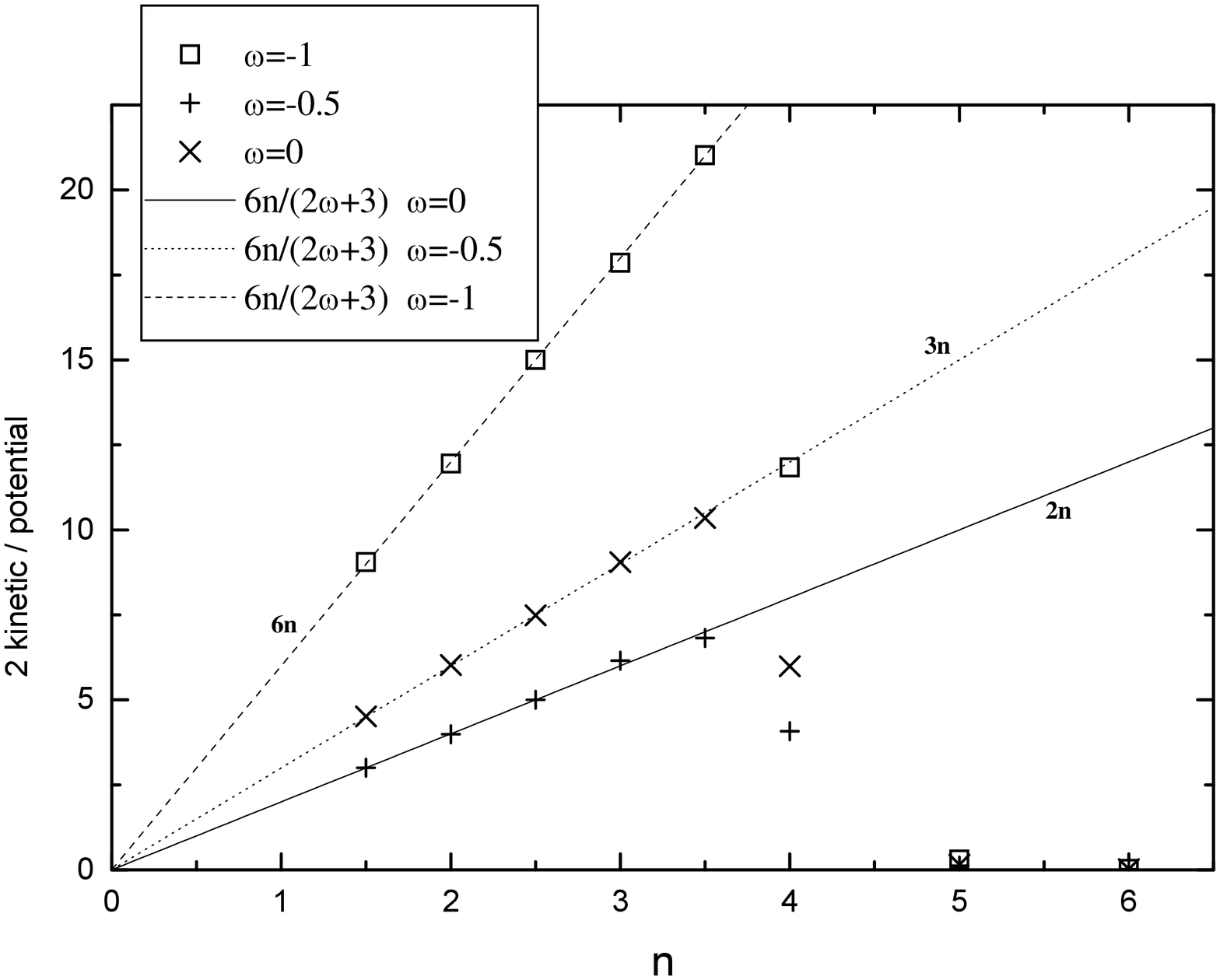}
\caption{The virial ratio ${{\bar{\dot{\delta\phi^2}}} \over
{\bar{\rho_v}\bar{\delta\phi^n}}}$ as a function of $n$ for three
values of $\omega$ ($\omega=0$ lower line, $\omega=-0.5$, and
$\omega=-1$ upper line). Notice that the virial theorem (straight
lines) is violated for $n\gsim 4$} \label{fig1}
\end{figure}
In Figure 1 we show the virial ratio ${{\bar{\dot{\delta\phi^2}}}
\over {\bar{\rho_v}\bar{\delta\phi^n}}}$ as a function of $n$ for
three values of $\omega$. The points correspond to numerically
obtained values while the lines are the corresponding analytical
results (\ref{virial}). For $n<4$ the analytically obtained
results are well verified by the numerics. For $n\geq 4$ however
the agreement is not good and the virial ratio rapidly drops
instead of increasing. This domination of the potential energy is
due to the violation of the condition (\ref{oscdom}). The
evolution of $\phi$ for $n=6$ shown in Figure 2 indicates that the
oscillations of $\delta \phi$ do not even overlap with $\delta\phi
=0$ and therefore the assumption $\bar{\delta\phi}=0$ used in
deriving the virial theorem is not realized for $n\geq 4$. This
may be understood by comparing the redshift rate of \be
{\delta\phi}_{min}\sim a^{-3/(n-1)}\ee (cf equation (\ref{dfmin}))
with that of \be \label{maxev} {\delta\phi}_{max}\sim
a^{-6/(n+2)}\ee which will be derived in the next section (see
equation (\ref{veffev1})). It is easy to see that for $n<4$,
${\delta\phi}_{max}$ eventually dominates while for $n>4$ it is
${\delta\phi}_{min}$ that dominates leading to violation of the
virial theorem and vanishing of the virial ratio since the
potential energy dominates.
\begin{figure}
\centering
\includegraphics[height=8cm,angle=-90]{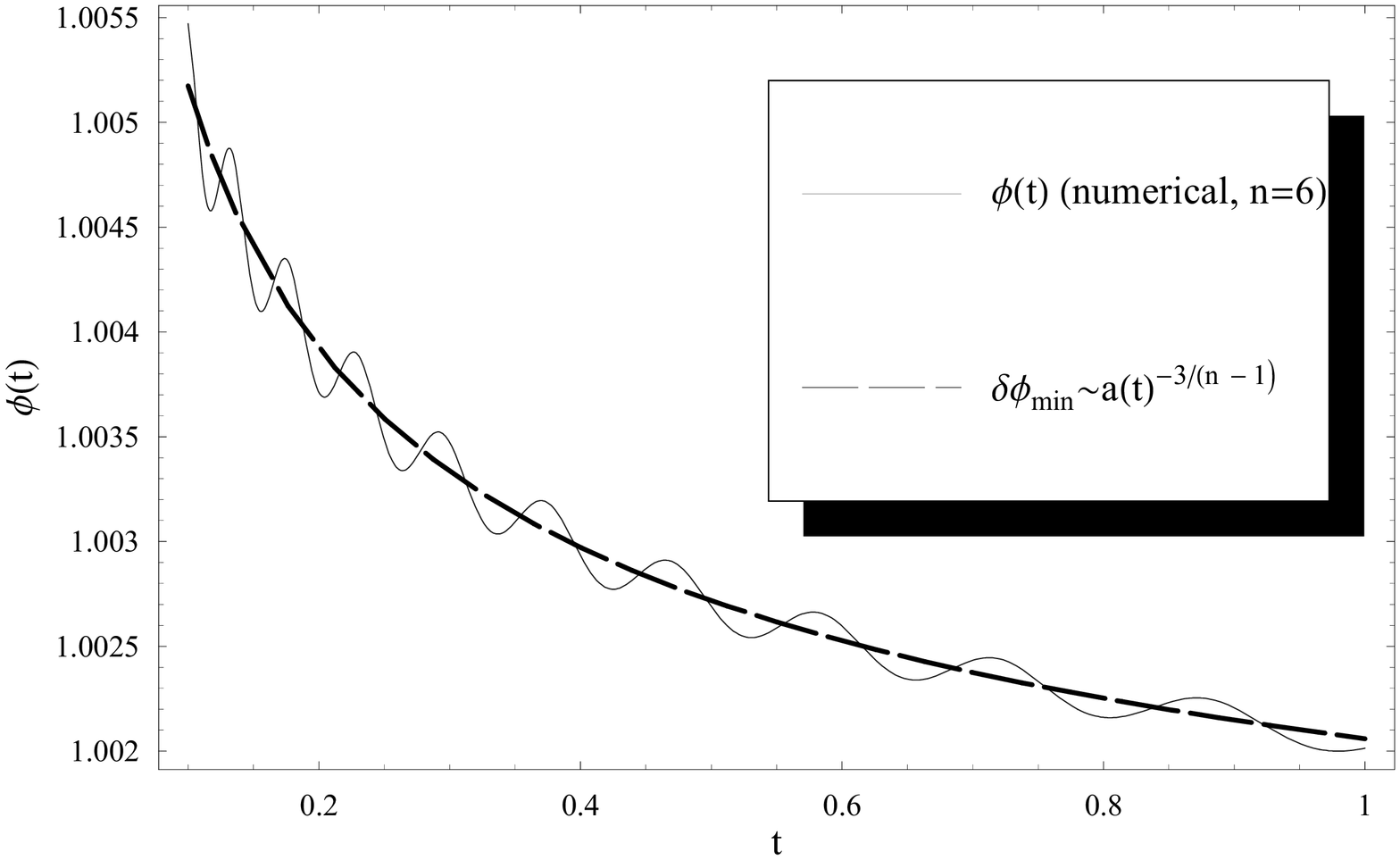}
\caption{The oscillations of $\delta \phi$ do not even overlap
with $\delta\phi =0$ for $n=6$. The equilibrium value
${\delta\phi}_{min}$ is larger than the oscillation amplitude
${\delta\phi}_{max}$ and redshifts like $a^{-3/5}\sim t^{-2/5}$
(matter energy dominated in the evolution). Thus the assumptions
of the virial theorem are violated.} \label{fig2}
\end{figure}
The fact that the oscillations of $\phi(t)$ are effectively around
$\phi =1$ for $n<4$ is demonstrated in Figure 3 which shows the
evolution of $\phi(t)$ for $n=1.5$. The oscillation amplitude
${\delta\phi}_{max}$ redshifts like $a^{-6/(n+2)}$ as predicted
analytically (see equation (\ref{veffev1})).
\begin{figure}
\centering
\includegraphics[height=8cm,angle=-90]{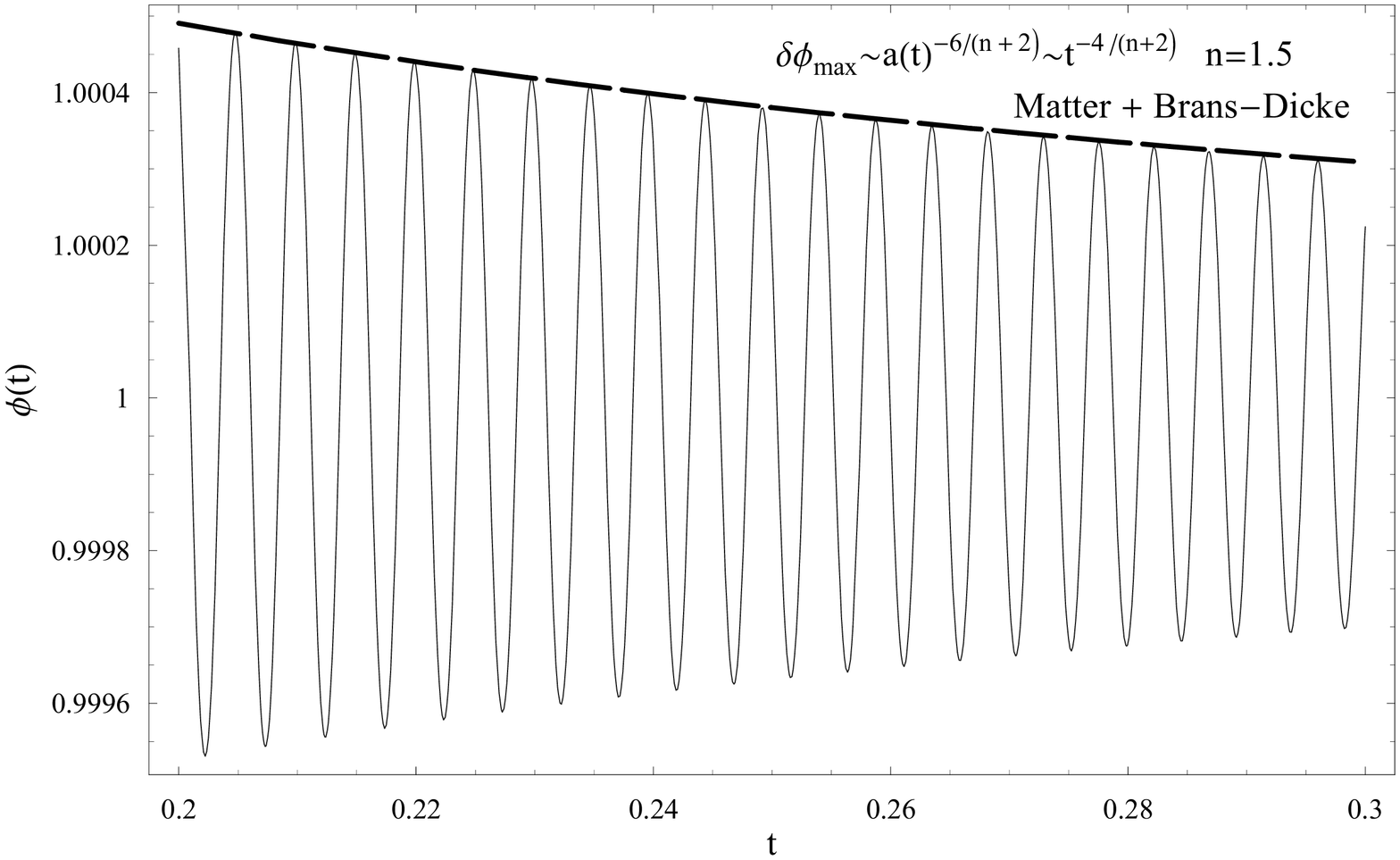}
\caption{For $n=1.5$ the BD scalar oscillations are approximately
symmetric around $\phi\simeq 1$ ($\omega=0$ was used in the
evolution but the result was found to be independent of $\omega$)}
\label{fig3}
\end{figure}
The period of the oscillations varies slowly with time and it can
be calculated using equation (\ref{bdecon1}) in a similar way as
for a minimally coupled scalar (equation (\ref{tmax})). The result
is that the oscillation period $T$ redshifts like \be \label{tred}
T \sim a^{{3(n-2)}\over {n+2}} \ee i.e. it decreases for $n<2$
while it increases for $n>2$. This result is confirmed numerically
in Figure 4 where we show $ln({{T(a)}\over {T(a=1)}})$ vs $ln(a)$.
The numerically obtained points are in good agreement with the
analytically predicted corresponding lines for $n=1.5$ (decreasing
T), $n=2$ (constant T) and $n=3.0$ (increasing T).
\begin{figure}
\centering
\includegraphics[height=8cm,angle=-90]{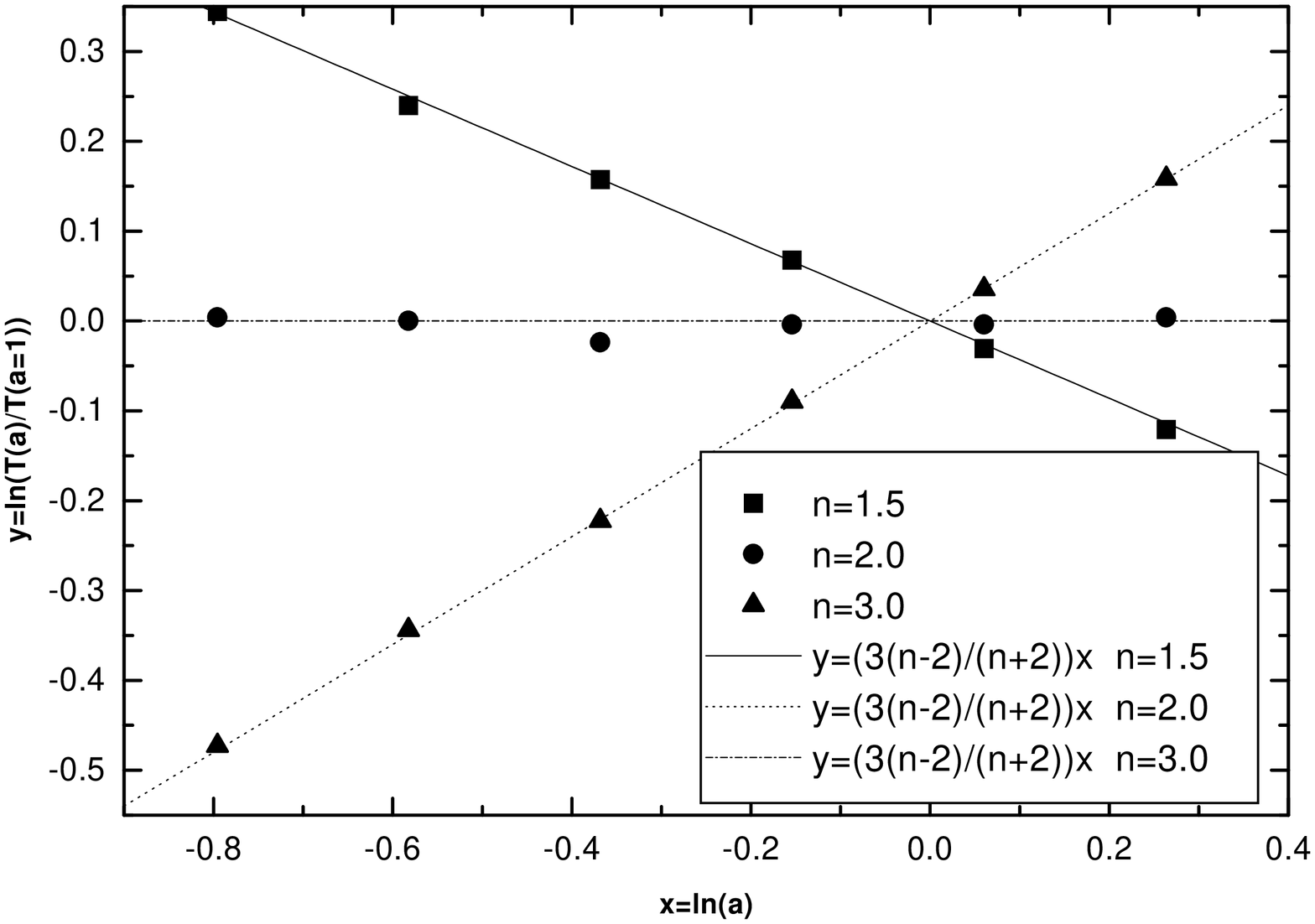}
\caption{The oscillation period increases (decreases) for $n>2$
($n<2$).} \label{fig4}
\end{figure}

We are now in position to calculate the equation of state that
connects the mean effective pressure with the mean effective
energy density of the BD oscillating field. Using the rescaling of
equations (\ref{tresc})-(\ref{rhoresc}) and equations
(\ref{rhoeq}), (\ref{peq}) we find that for small approximately
symmetric oscillations \ba \label{rhoeff1} \bar{\rho}_\phi &\simeq
& {\omega \over 6} \bar{\dot{\delta \phi^2}}  + \bar{\rho}_v
\bar{\delta \phi^n} \\
\bar{p}_\phi &\simeq & {\omega \over 6} \bar{\dot{\delta \phi^2}}
- \bar{\rho}_v \bar{\delta \phi^n} \ea Thus keeping the notation
of section II we have \be  \gamma = {{\bar{\rho}_\phi +
\bar{p}_\phi}\over {\bar{\rho}_\phi}} = {{{\omega \over
3}\bar{\dot{\delta \phi^2}}} \over {\bar{\rho}_\phi}} \ee Using
now equations (\ref{intrat2}) - (\ref{virial}) we find  \be
\label{gamma3} \gamma = {{2 \omega n}\over {\omega n + 2 \omega +
3}} = w + 1 \ee or \be \label{wpar} w= {{\omega n - (2\omega +
3)}\over {\omega n + (2\omega + 3)}} \ee For a minimally coupled
scalar this result would immediately imply that the redshift of
the energy density is of the form of equation (\ref{rhoa}) with
$\gamma$ given by (\ref{gamma3}). If that were the case,
superacceleration\cite{Torres:2002pe,Faraoni:2001tq} ($\dot{H}>0$
which may be favored observationally\cite{Caldwell:1999ew}) would
be easily obtained for negative values of $\omega$ which lead to
$w<-1$ (see Figure 5). For a BD scalar however this is not the
case. It is easily seen by using e.g. equation (\ref{bdecon}) that
the continuity equation \be \label{conteq} \dot{\bar{\rho_\phi}} +
3 {\dot{a}\over a}({\bar{\rho}_\phi + \bar{p}_\phi})=0 \ee is not
realized\cite{Torres:2002pe} for $\bar{\rho}_\phi$ and
$\bar{p}_\phi$.
\begin{figure}
\centering
\includegraphics[height=11cm,angle=-90]{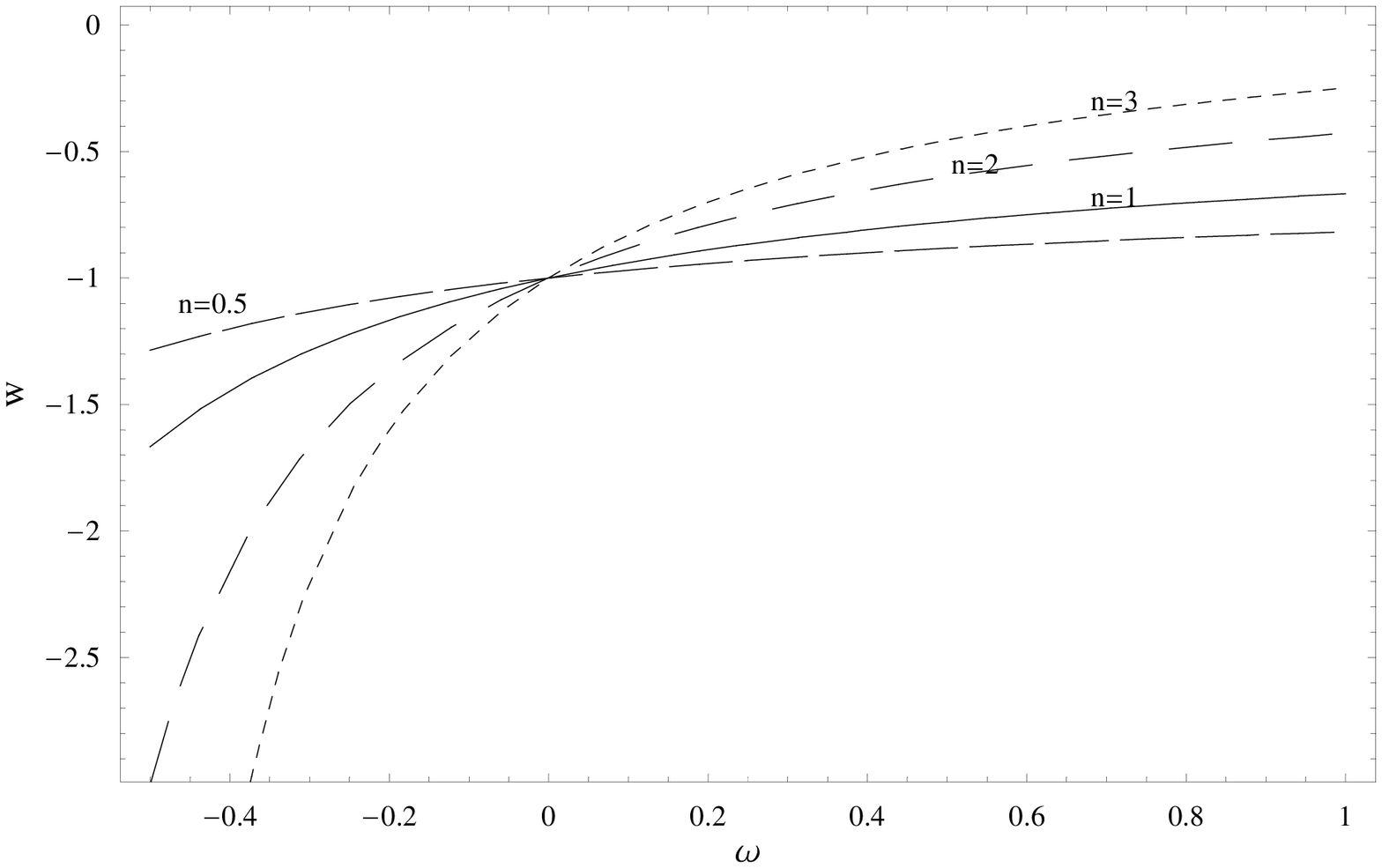}
\caption{The equation of state parameter $w$ is less than $-1$ for
negative $\omega$.} \label{fig5}
\end{figure}
Therefore $\bar{\rho}_\phi$ does not redshift like $a^{-3
\gamma}$. The redshift rate of the effective energy density and
the corresponding evolution of the scale factor $a(t)$ are studied
in the next section.

\section{Energy Redshift and Scale Factor Evolution}

Even though the continuity equation (\ref{conteq}) is not realized
for a BD scalar, equation (\ref{bdecon}) can be used to derive the
effective energy density redshift. For symmetric oscillations this
equation implies \be \label{effecon} {d\over {dt}}
({{\bar{\dot{\delta \phi^2}}}\over 2} +{{6 \bar{\rho_v}}\over {2
\omega + 3}} \bar{\delta \phi^n}) = - 3{\dot
a\over{a}}\bar{\dot{\delta\phi^2}} \ee Using now equations
(\ref{bdecon1}) and (\ref{intrat2}) we find \be {d \over {dt}}
V_{max}^{eff} = - 6 {\dot a\over{a}} V_{max}^{eff} {n\over {n+2}}
\ee which implies \be \label{veffev1} V_{max}^{eff} \sim
a^{-{{6n}\over{n+2}}} \sim \delta\phi_{max}^n \ee and leads to
equation (\ref{maxev}) used in the previous section. Using now
equations (\ref{rhoeff1}), (\ref{intrat2}), (\ref{veff1}) and
(\ref{veffev1}) we find \be \label{rhoev2} \bar{\rho}_\phi \sim
V_{max}^{eff}\sim a^{-{{6n}\over {n+2}}}\ee which gives the
redshift rate of the BD energy density. It is interesting to note
that even though the equation of state parameter $w$ differs from
that of a minimally coupled scalar and depends on $\omega$, the
redshift rate of the energy density is independent of $\omega$ and
is similar to the corresponding result of a minimally coupled
scalar. This result is numerically verified in Figure 6 where we
show a plot of $-ln(\rho_\phi (a)/\rho_\phi (a=1))$ vs $ln(a)$
obtained numerically for various values of the scale factor $a$
and for three values of $n$ ($n=1.5$, $n=2.0$ and $n=3$).
According to equation (\ref{rhoev2}), the numerically obtained
points should lie on straight lines with slope $3\Gamma$ where \be
\label{bigg} \Gamma = {{2n}\over {n+2}} \ee This is indeed
realized as shown in Figure 6 where we also show these straight
lines.
\begin{figure}
\centering
\includegraphics[height=8cm,angle=-90]{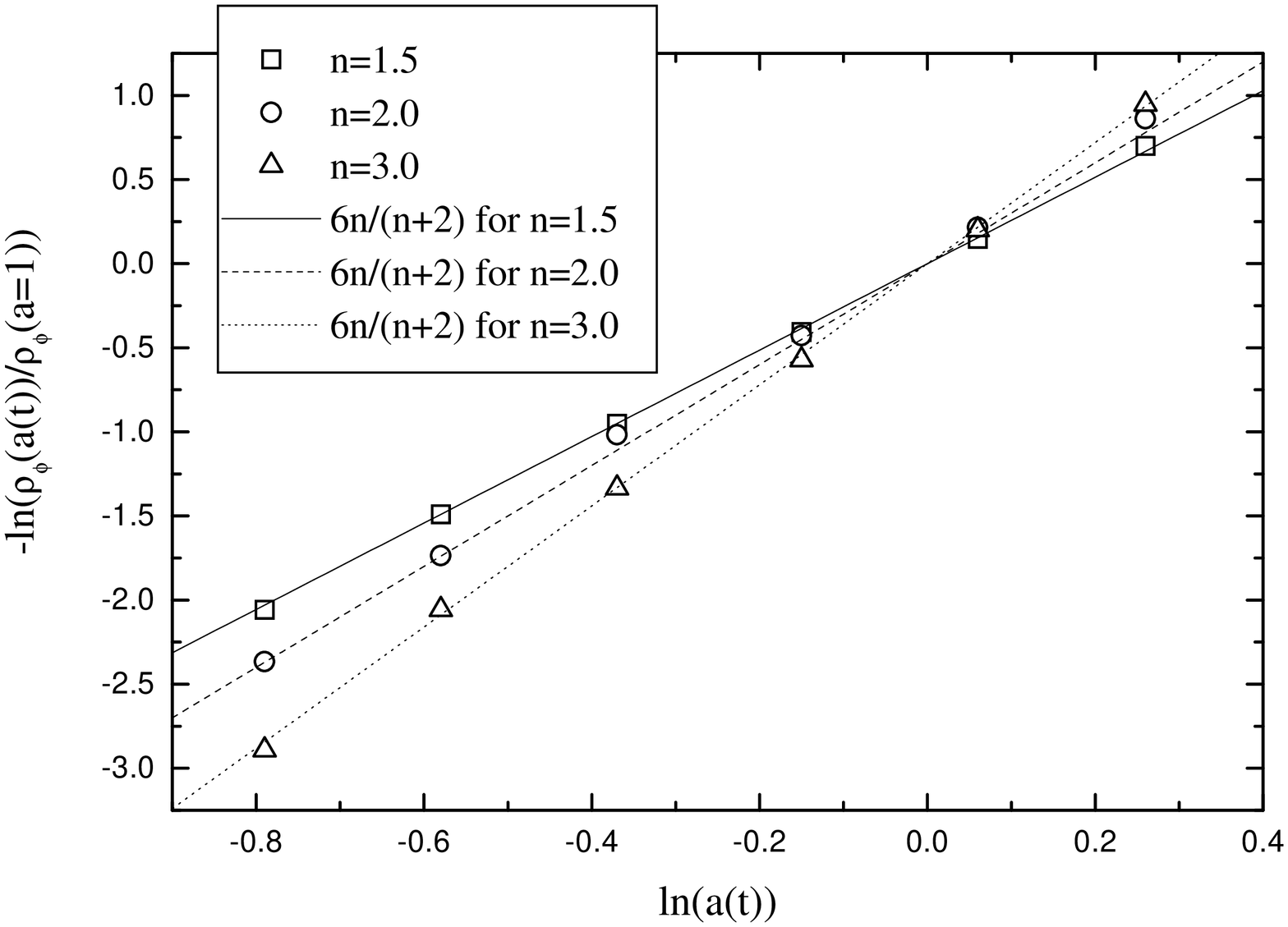}
\caption{The energy density of the oscillating BD scalar redshifts
according to the analytical predictions for three values of $n$.}
\label{fig6}
\end{figure}

We now focus on the evolution of the scale factor in the presence
of an oscillating BD scalar given the above derived energy density
redshift. To isolate the effects of the scalar we ignore the
presence of matter at this stage. Thus the rescaled Friedman
equation (\ref{bdd1}) is approximated by
\begin{equation} \label{bddd1}
 {\dot a^2\over{a^2}}\simeq {\omega\over {6}}
\bar{\dot{\delta \phi^2}}+ \bar{\rho_v}\bar{\delta\phi^n}\simeq B
a^{-3\Gamma}
\end{equation}
where B is a constant and $\Gamma$ determines the redshift rate of
$\bar{\rho_\phi}$ and is given by equation (\ref{bigg}). This
implies that \be \label{aevol3} a\sim t^{{2\over {3\Gamma}}} \sim
t^{{n+2}\over {3n}} \ee Thus even though $\gamma \neq \Gamma$ in
the case of a BD scalar oscillations, the induced scale factor
evolution is identical to that of a minimally coupled scalar shown
in equation (\ref{aevol1}).  This result is numerically verified
in Figure 7 where we show a plot of $ln({{a(t)}\over {a_0}})$ vs
$ln({t\over t_0})$ (we chose $t_0 \simeq 2/3$ and $a_0 =a(t_0)$).
The dark continous line corresponds to the numerical scale factor
evolution for $n=1.5$ while the light dashed line is the
corresponding analytical result of equation (\ref{aevol3}).
\begin{figure}
\centering
\includegraphics[height=9cm,angle=-90]{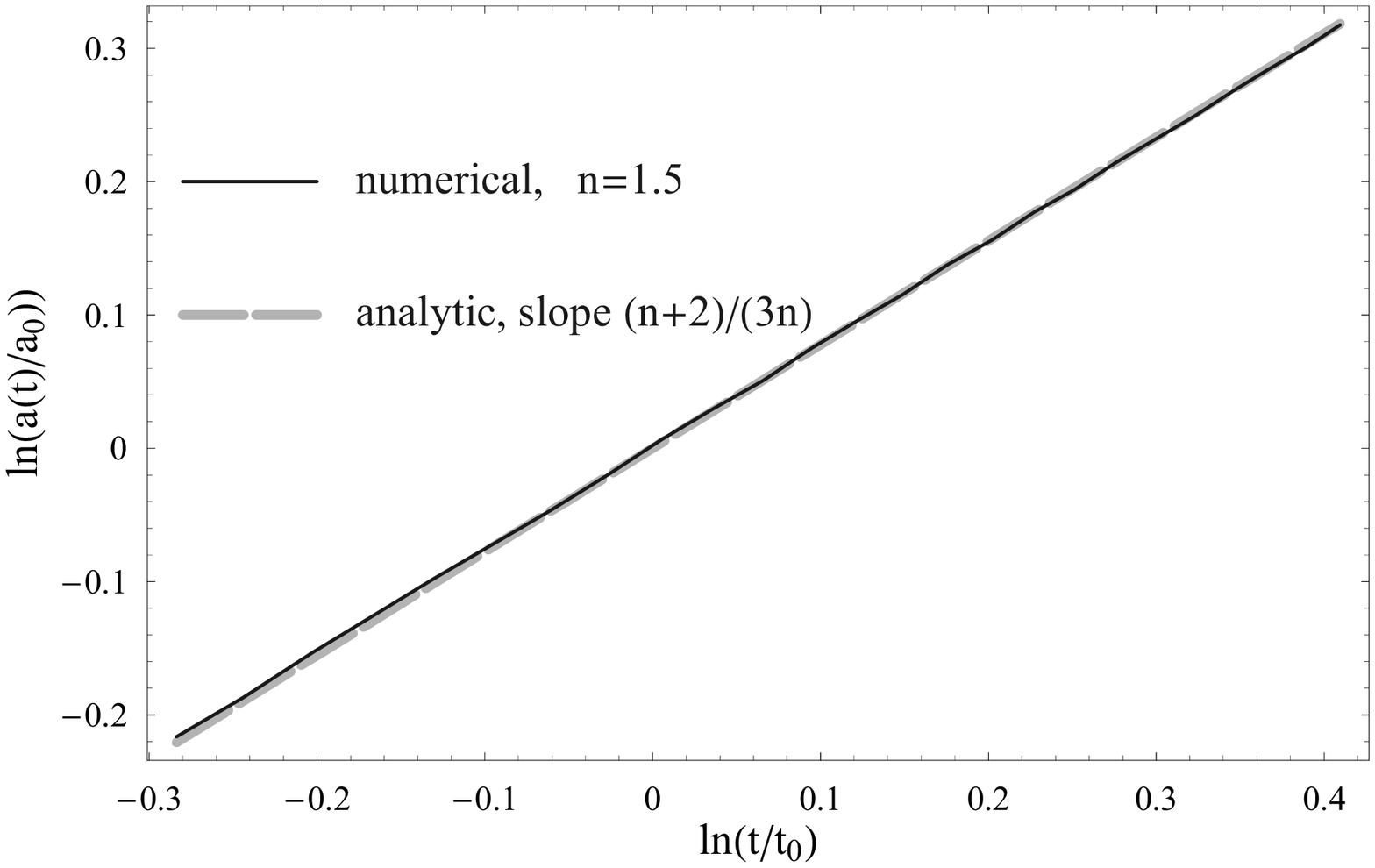}
\caption{The numerically obtained evolution of the scale factor is
in good agreement with the analytical prediction for $n=1.5$ when
matter is not present.} \label{fig7}
\end{figure}
The agreement is quite good, confirming the analytical
expectation. It is interesting to note that this result implies
accelerated expansion for $0<n<1$ as in the case of a minimally
coupled scalar. The oscillations in this case however can be
generically induced not just by initial conditions but also by the
time dependent redshift of the minimum $\delta\phi_{eql}$ of the
effective potential (\ref{veff0}).

The scale factor evolution derived above when matter is not
present can be combined with equation (\ref{maxev}) to find the
time evolution of $\delta \phi_{max}$. The predicted power law
$\delta\phi_{max} \sim t^{2/n}$ is verified numerically in Figure
8 for $n=1.5$.
\begin{figure}
\centering
\includegraphics[height=9cm,angle=-90]{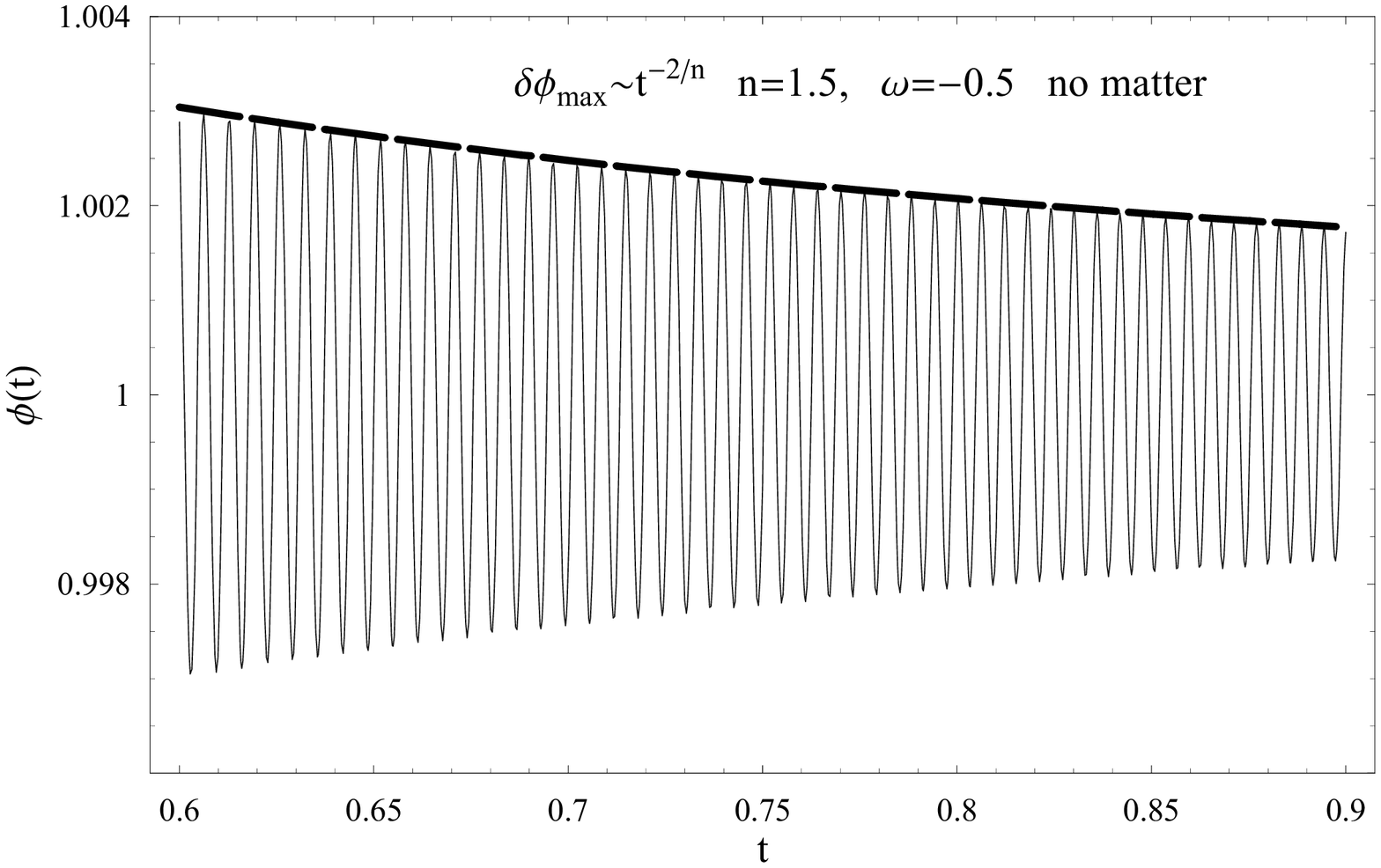}
\caption{The numerically obtained evolution of the
$\delta\phi_{max}$ is in good agreement with the analytical
prediction $\delta\phi_{max}\sim t^{2/n}$for $n=1.5$.}
\label{fig8}
\end{figure}
\section{Radion Oscillations - Outlook}
The above derived results can easily be applied to the case of
radion dynamics. By setting $\omega=-1+1/D$ in equation
(\ref{wpar}) we can obtain the equation of state parameter for
high frequency radion oscillations in terms of the number of extra
dimensions $D$ as \be \label{wrad} w ={{n (D-1)+(D+2)}\over {n
(D-1)-(D+2)}} \ee Clearly $w$ is negative and in fact less than
$-1$ for a wide range of parameters $n$, $D$. The evolution of the
scale factor however is independent of $D$ and is given by
equation (\ref{aevol3}). This result is in agreement with Ref.
\cite{Perivolaropoulos:2002pn} where the case $n=2$ was studied.

Even though radion oscillations can not lead to superacceleration,
they do lead to accelerating expansion for $n<1$ as in the case of
a minimally coupled scalar field. Thus, the mechanism of Ref.
\cite{Sahni:1999qe} of quintessence from an oscillating minimally
coupled scalar with $n<1$ can be extended to the case of an
oscillating BD scalar and in particular the case of an oscillating
radion. In fact an oscillating radion has two potential advantages
over an oscillating minimally coupled scalar. First its origin is
better defined and better motivated since it is a generic
geometrical prediction of theories with extra dimensions. Second,
as discussed below, tracking can in principle be achieved at early
times by the coupling to the matter energy momentum tensor which
generically induces radion oscillations due to the matter redshift
(the effective potential is time dependent).  Since this
possibility does not exist for a minimally coupled scalar, an
exponential potential has to be used\cite{Sahni:1999qe} for large
$\phi$ so that the energy density in $\phi$ tracks the matter
component at early times while at late times the field oscillates
around its potential minimum with $n<1$ thus producing the
observed accelerated evolution of the scale factor. The potential
used in Ref. \cite{Sahni:1999qe} that combines an exponential
behavior at large $\phi$ with power law behavior with power index
$n$ around its minimum is \be V(\phi) = V_0 (cosh\; \lambda \phi
-1)^{n/2} \ee Our results on the scale factor evolution indicate
that the analysis of Ref. \cite{Sahni:1999qe} is applicable to the
case of radion oscillations even though the equation of state is
different in the case of the radion. However, the coupling of the
radion to the trace of the matter energy momentum tensor opens up
a new possibility for achieving tracking at early times. The
radion could start very close to its effective potential $\delta
\phi_{eql}$ minimum at early times with subdominant oscillations
($\delta \phi_{max}\ll \delta \phi_{eql}$) induced by the time
dependent shift of the minimum of the effective potential
(\ref{veff0}). The radion energy during this early evolution can
track the matter energy density. At late times when $\delta
\phi_{max}>\delta \phi_{eql}$ the oscillations can dominate and
lead to accelerating expansion for $n<1$. This mechanism has the
possible advantage of avoiding the introduction of a tuned
potential. Instead only the existence of a stabilizing potential
minimum is required and a power law behavior around it. The
viability of this type of mechanism is an interesting problem
worth of further investigation.

A possible direct observational signature of radion oscillations
is the existence of a coherent high frequency gravitational wave
background. Even though the frequency of these gravitational waves
is expected to be of order  $10^{12}Hz$ and is much beyond the
range of current experiments, a detailed study of their features
could be interesting. A study of gravitational wave backgrounds
from extra dimensions induced by other mechanisms can be found in
Ref. \cite{Hogan:2000is}.

The stability of the field configurations against spatial
fluctuations is also an interesting issue. Homogeneous oscillating
scalar fields have been shown to fragment into baryons
(Affleck-Dine mechanism\cite{Affleck:1984fy}), Q-Balls
\cite{Kasuya:1999wu} or I-Balls \cite{Kasuya:2002zs} depending on
their potential and number of components. Since the oscillating
radion corresponds to a real scalar field it may be possible to
fragment into a generalization of I-Balls (BD I-Balls) with
interesting cosmological consequences.

{\bf Acknowledgements:} I thank G. Leontaris, I. Rizos and T.
Vachaspati for useful discussions. This work was supported by the
European Research and Training Network HPRN-CT-2000-00152.

\end{document}